\documentclass[12pt,flequ]{article}
\usepackage{amsmath,amsthm,amssymb,natbib,color}
\usepackage{graphicx}
\newtheorem{theorem}{Theorem}
\newtheorem{remark}{Remark}

\titlepage
  \def\lg{\large}  

\begin{document}

\numberwithin{equation}{section}

\begin{center}
{\lg \bf Bias-corrected GEE estimation and smooth-threshold GEE variable selection for single-index models with clustered data}
\end{center}

 \vskip 0.8 true cm

 \begin{center}
 Peng Lai$^{a,d}$, Qihua Wang$^{b,c}$ and  Heng Lian$^{d}$ \\

{\small
 {\it $^{a}$College of Math $\&$ Physics, Nanjing University of Information Science $\&$ Technology, \\Nanjing 210044, China}\\
 {\it $^b$Academy of Mathematics and Systems Science, Chinese Academy of Science, \\Beijing 100190, China}\\
 {\it $^{c}$School of Mathematics and Statistics, Yunnan University, \\Kunming 650091, China}\\
and\\
{\it $^{d}$Division of Mathematical Sciences, School of Physical and Mathematical Sciences\\ Nanyang Technological University, Singapore 637371}\\

 Email: henglian@ntu.edu.sg}

\end{center}

\thispagestyle{empty}

\vskip 1.0 cm

\begin{abstract}
In this paper, we present a generalized estimating equations based estimation approach and a variable selection procedure for single-index models when the observed data are clustered. Unlike the case of independent observations, bias-correction is necessary when general working correlation matrices are used in the estimating equations. Our variable selection procedure based on smooth-threshold estimating equations \citep{Ueki-2009} can automatically eliminate irrelevant parameters by setting them as zeros and is computationally simpler than alternative approaches based on shrinkage penalty. The resulting estimator consistently identifies the significant variables in the index, even when the working correlation matrix is misspecified. The asymptotic property of the estimator is the same whether or not the nonzero parameters are known (in both cases we use the same estimating equations), thus achieving the oracle property in the sense of \cite{Fan-Li-2001}. The finite sample properties of the estimator are illustrated by some simulation examples, as well as a real data application.

\end{abstract}

\noindent{\bf Keywords and phrases:} Generalized estimating equation; Longitudinal
data; Oracle property; Single-index model; Variable selection.

\noindent{\bf Short title:} Variable selection for SIM

\newpage
\baselineskip=0.7 true cm

\section{Introduction}
Many data sets nowadays are characterized by two properties that
make their statistical analysis complicated, high-dimensionality and
dependence of observations. In fact, clustered data with a medium to
large number of covariates are often produced in fields such as
biology, engineering, or medicine. For different clusters $1\le i\le
n$, let $Y_i=(Y_{i1},\ldots,Y_{im_i})^T$ denote the vector of
outcome values, which depends on a $p\times m_i$ covariate matrix
$\mathbb{X}_i=(X_{i1},\ldots,X_{im_i})$,
$X_{ij}=(X_{ij1},\ldots,X_{ijp})^T$. When the dimension of $X_{ij}$ is high, it is worthwhile to spend efforts in seeking a more parsimonious representation of the regression function in the hope of making estimation feasible for moderate sample size. Dimension reduction is one way towards this goal. As a popular instantiation of dimensional reduction idea, the single-index model for the
clustered data is defined by
\begin{equation}
Y_i=\mathbf{g}(\mathbb{X}_i^{\top}\beta)+\varepsilon_i,
i=1,2,\ldots,n, \label{1.1}
\end{equation}
where
$$
\mathbf{g}(\mathbb{X}^{\top}_i\beta)=\left(
      \begin{array}{ccc}
        g(X_{i1}^{\top}\beta) \\
        \vdots  \\
        g(X_{im_i}^{\top}\beta) \\
      \end{array}
    \right),
\varepsilon_i=\left(
             \begin{array}{c}
               \varepsilon_{i1} \\
               \vdots \\
               \varepsilon_{im_i} \\
             \end{array}
           \right).
$$
Here $g(\cdot)$ is an unknown link function and $\varepsilon_i$ is
mean-zero random error with covariance matrix
$Var(\varepsilon_i)=\Sigma_i$ for the $ith$ subject, and
$\beta=(\beta_1,\ldots,\beta_p)^T$ is the unknown parameters for the
index associated with covariates. Since both $g$ and $\beta$ are
unknown, it is commonly assumed that $\parallel\beta\parallel=1$ for
identifiability, where $\parallel\cdot\parallel$ is the Euclidean
norm. The true value of $\beta$ will be denoted by $\beta_0$.
Throughout this paper we assume that the total sample size
$N=\sum_{i=1}^n m_i$ is large (diverges to $\infty$ in our
theoretical investigations) while $\{m_i,i=1,\ldots,n\}$ are
uniformly bounded.

The popularity of the semiparametric single-index model presented
above can be attributed to its ability to address the so-called
``curse of dimensionality" problem in multi-dimensional
nonparametric regression by making use of a combination of
predictors as univariate index, which hopefully can still capture
some important relationships between the covariates and the
responses. As a dimension reduction method, single-index models have
been studied extensively. See for example, \cite{Ichimura-1993,
Hardle-Hall-Ichimura-1993, Carroll-Fan-Gijbels-Wand-1997,
Xia-Tong-Li-1999, Naik-Tsai-2000, Naik-Tsai-2001, Naik-Tsai-2004,
Yu-Ruppert-2002,Delecroix-Hardle-Hristache-2003, Zhu-Xue-2006,
Xia-Hardle-2006, Kong-Xia-2007, Wong-Ip-Zhang-2008}. More recently,
\cite{Bai-Fung-Zhu-2009} studied the single-index model for
longitudinal data, and proposed to use splines to estimate $\beta$
and the unknown link function based on quadratic inference
functions. Our study here is different from that work in many
respects. \cite{Bai-Fung-Zhu-2009} considered asymptotic analysis
with a fixed number of knots and thus their analysis is not
appropriate when the true link function is not inside the spline
space. In particular, their asymptotic analysis is only for a parametric model since the number of 
unknown parameters does not diverge with sample size. Our estimation method and asymptotic analysis does not pose this constraint, and treat the unknown link function as a truly nonparametric component. Furthermore, we will consider variable
selection problem which was not investigated before for single-index models on longitudinal data.

Even though single-index models avoid the problem of ``curse of
dimensionality" to some extent, in practice, one would still want to
investigate which covariates are relevant for prediction, both for
better interpretation of the model, and for better efficiency of the
estimator. In recent years, penalization or shrinkage based variable
selection methods have attracted lots of attention, due to their
computational efficiency for high-dimensional problems, and their
statistical stability compared to information criterion based
methods \citep{Fan-Li-2001,Zou-2006}.
Examples of shrinkage estimation methods include LASSO \citep{Tibshirani-1996}, SCAD \citep{Fan-Li-2001}, Adaptive Lasso \citep{Zou-2006}, Dantzig selector \citep{Candes-Tao-2007}, and many others. For single-index models, \cite{Naik-Tsai-2001} considers variable selection using sliced inverse regression, \cite{Kong-Xia-2007} uses cross-validation to select the significant variables, but these estimators are not expected to have the oracle property \citep{Fan-Li-2001}. %Zhu and Zhu (2009)$^{\cite{Zhu-Zhu-2009}}$ combines the sufficient dimension reduction method with SCAD penalty method to do variable selection.
%We'd like to study the variable selection problem for single-index model for clustered data to improve the estimating performance, and study its oracle property.

In this paper, we build on the estimating equations based approach for single-index models \citep{chang10}, which was shown to result in a more efficient estimator for the index vector, and extend it to the case where data are clustered. \textcolor{black}{The bias-corrected estimating equations we use here were proposed in \cite{Li-Zhu-Xue-Feng-2010}, which focused on the
construction of confidence regions of partially linear single-index models
for longitudinal data through the empirical likelihood method.}
Furthermore, variable selection is achieved by extending the smooth-threshold estimating equations proposed in \cite{Ueki-2009}. Compared to shrinkage methods reviewed above, this approach dispenses with convex optimization and is thus computationally simpler. We will theoretically demonstrate the oracle property of the estimator as well as empirically illustrate its performance. \textcolor{black}{We also note that recently \cite{cui-2011} has extended the estimating equations approach to generalized single-index models which do not involve clustered data. We expect that this can also be extended to the case with variable selection for clustered data, although this is outside the scope of the current paper.}

The rest of the paper is organized as follows. In Section 2 we present our estimation approach for single-index models with clustered data, and in Section 3 a variable selection procedure based on smooth-threshold generalized estimating equations is presented. The oracle property for the proposed estimator is also discussed. In Section 4, we report some simulation studies as well as an application to a real data set. Our simulations show the advantage of incorporating the intra-cluster correlation in estimation. The proofs of theoretic results are presented in the Appendix.

\section{Bias-corrected GEE estimation}
In model (\ref{1.1}), we imposed  $\|\beta\|=1$ for identifiability, which implies that the parameter is not an interior point of the $p$-dimensional space, causing some difficulty in inference. We use the ``remove one component" method used previously in \cite{Yu-Ruppert-2002, Zhu-Xue-2006, chang10}. Without loss of generality, we
assume that for some $1\le r\le p$, $\beta_r>0$. Let $\beta^{(r)}=(\beta_1,\ldots,\beta_{r-1},\beta_{r+1},\ldots,\beta_p)^T$ be the $(p-1)$-dimensional parameter vector after removing the $r$th component $\beta_r$ of $\beta$. Then, we may write
$$
\beta(\beta^{(r)})=(\beta_1,\ldots,\beta_{r-1},(1-\|\beta^{(r)}\|^2)^{1/2},\beta_{r+1},\ldots,\beta_p)^T.
$$
Since $\|\beta_0^{(r)}\|<1$, $\beta(\cdot)$ is infinitely differentiable in a neighborhood of $\beta_0^{(r)}$, and the Jacobian is
\begin{equation*}
J_{\beta^{(r)}}=\frac{\partial\beta}{\partial\beta^{(r)}}=(b_1,\ldots,b_p)^T,\label{2.1}
\end{equation*}
where $b_s$ is a $(p-1)$-dimensional unit vector with $s$th component 1 for $s\neq r$, and $b_r=-(1-\|\beta^{(r)}\|^2)^{-1/2}\beta^{(r)}$.

Based on these notations, we construct the generalized estimating equation (GEE) for the single-index model with clustered data as %For $\beta\in\mathbf{B}=\{\beta:\|\beta-\beta_0\|\leq C_1n^{-1/2}\}$, or $\beta^{(r)}\in\mathbf{B}_n=\{\beta^{(r)}:\|\beta^{(r)}-\beta_0^{(r)}\|\leq C_2n^{-1/2}\}$, the estimating equation can be constructed as
\begin{equation}
\sum_{j=1}^nZ_j^TR_j^{-1}(Y_j-\mathbf{g}(\mathbb{X}_j^T\beta))=0,
\label{2.2}
\end{equation}
where
$$
Z_j=\left(
      \begin{array}{c}
        g'(X_{j1}^T\beta)(J_{\beta^{(r)}}^TX_{j1})^T \\
        \vdots \\
        g'(X_{jm_j}^T\beta)(J_{\beta^{(r)}}^TX_{jm_j})^T \\
      \end{array}
    \right),
    j=1,\ldots,n,
$$
and $R_j, j=1,\ldots,n$ are the working covariance matrices, possibly depending on some unknown parameter $\alpha$, which can be estimated by the method of \cite{Liang-Zeger-1986}. From the estimating equations, we can see that if $R_j=I_{m_j}$, with $I_{m_j}$ the $m_j\times m_j$ identity matrix, we just ignore the dependence of the data within a cluster, that is, assume working independence \citep{Lin-Carroll-2000}. For the following theoretical results, we do not require $R_j$ to be the same as the true covariance $\Sigma_j$, although $R_j=\Sigma_j$ results in the most efficient estimator.

The estimating equation (\ref{2.2}) contains the unknown functions $g(\cdot)$ and $g'(\cdot)$. To solve this problem, we need to plug in some estimates for these two unknown functions. Here we use the local linear regression \citep{Fan-Gijbels-1996}. Similar to \cite{chang10}, for any given $\beta$, we can estimate $g(t)$ and $g'(t)$ by minimizing
\begin{equation*}
\min_{a,
b}{\sum_{i=1}^n\sum_{j=1}^{m_i}\{Y_{ij}-a-b(X_{ij}^T\beta-t)\}^2K_h(X_{ij}^T\beta-t)},\label{2.3}
\end{equation*}
where $K$ is a kernel function, $K_h(\cdot)=K(\cdot/h)/h$ and $h$ is the bandwidth. Let $(\hat{a}, \hat{b})$ be the minimizers and set $\hat{g}(t,\beta)=\hat{a}$ and $\hat{g}'(t,\beta)=\hat{b}$. Simple and standard calculations yield the closed form expression
\begin{equation}
\hat{g}(t,\beta)=\sum_{i=1}^n\sum_{j=1}^{m_i}W_{nij}(t,\beta)Y_{ij}\quad
and \quad
\hat{g}'(t,\beta)=\sum_{i=1}^n\sum_{j=1}^{m_i}\tilde{W}_{nij}(t,\beta)Y_{ij},\label{2.4}
\end{equation}
where
$$
W_{nij}(t,\beta)=\frac{U_{nij}(t,\beta)}{\sum_{i=1}^n\sum_{j=1}^{m_i}U_{nij}(t,\beta)},
\quad
\tilde{W}_{nij}(t,\beta)=\frac{\tilde{U}_{nij}(t,\beta)}{\sum_{i=1}^n\sum_{j=1}^{m_i}U_{nij}(t,\beta)},
$$
$$
U_{nij}(t,\beta)=K_h(X_{ij}^T\beta-t)\{S_{n,2}(t,\beta)-(X_{ij}^T\beta-t)S_{n,1}(t,\beta)\},
$$
$$
\tilde{U}_{nij}(t,\beta)=K_{h}(X_{ij}^T\beta-t)\{(X_{ij}^T\beta-t)S_{n,0}(t,\beta)-S_{n,1}(t,\beta)\},
$$
and
$$
S_{n,l}(t,\beta)=\frac{1}{N}\sum_{i=1}^n\sum_{j=1}^{m_i}(X_{ij}^T\beta-t)^lK_h(X_{ij}^T\beta-t),\quad
l=0,1,2.
$$
Plugging these estimators into (\ref{2.2}), we obtain the estimating equations
\begin{equation}
\sum_{j=1}^n\hat{Z}_j^TR_j^{-1}(Y_j-\hat{\mathbf{g}}(\mathbb{X}_j^T\beta))=0,
\label{2.5}
\end{equation}
where
$$
\hat{Z}_j=\left(
      \begin{array}{c}
        \hat{g}'(X_{j1}^T\beta)(J_{\beta^{(r)}}^TX_{j1})^T \\
        \vdots \\
        \hat{g}'(X_{jm_j}^T\beta)(J_{\beta^{(r)}}^TX_{jm_j})^T \\
      \end{array}
    \right),
    j=1,\ldots,n.
$$
We can also obtain an initial estimator of $\beta$, denoted by $\tilde{\beta}$, by assuming working independence. When assuming working independence, the results in \cite{Wang-Xue-Zhu-Chong-2010} \textcolor{black}{apply with few changes}, and in particular, $\tilde{\beta}$ is $\sqrt{n}$-consistent under standard assumptions.

For our theoretical analysis, we will assume that $R_1, \ldots,R_n$ are prespecified and known. We briefly discuss the more general case where $R_i$ must be estimated in Remark 1 below. 
\textcolor{black}{ However, when we do not assume that $R_j,j=1,\ldots,n,$ are all equal, similar to \cite{Wang-Xue-Zhu-Chong-2010},
(\ref{2.5}) leads to
\begin{eqnarray*}
&&\sum_{j=1}^n\hat{Z}_j^TR_j^{-1}(Y_j-\hat{\mathbf{g}}(\mathbb{X}_j^T\beta))\\
&&=U_1(\beta_0^{(r)})-nV_1(\tilde{\beta}^{(r)}-\beta_0^{(r)})-U_2(\tilde{\beta}^{(r)})+o_p(\sqrt{n}),
\end{eqnarray*}
where
\begin{eqnarray*}
U_1(\beta_0^{(r)})&=&\sum_{k=1}^n\big(\mathbf{g}'(\mathbb{X}_k^T{\beta}_0)
J_{{\beta}_0^{(r)}}^T(\mathbb{X}_k-E[\mathbb{X}_k|\mathbb{X}_k^T{\beta}_0])\big)^TR_k^{-1}\varepsilon_k\\
V_1&=&\lim_n\frac{1}{n}\sum_{k=1}^nE\Big[(\mathbf{g}'(\mathbb{X}_k^T{\beta}_0)J_{{\beta}_0^{(r)}}^T
\mathbb{X}_k)^TR_k^{-1}
(\mathbf{g}'({\mathbb{X}_k}^T{\beta}_0)J_{{\beta}_0^{(r)}}^T\mathbb{X}_k)\Big]
\end{eqnarray*}
and
\begin{eqnarray*}
U_{2s}(\beta_0^{(r)})&=&\sum_{k=1}^n\sum_{j=1}^{m_k}\sum_{i=1}^{m_k}\Big[\sum_{l_1=1}^n\sum_{l_2=1}^{m_{l_1}}
W_{nkj}(X_{l_1l_2}^T\beta_0,\beta_0)g'(X_{l_1i}^T\beta_0)X_{l_1is}^JR_{l1}^{il_2}\\
&-&g'(X_{ki}^T\beta_0)
E(X_{kis}^J|X_{ki}^T\beta_0)R_k^{ij}\Big]\varepsilon_{kj},
\end{eqnarray*}
$U_{2s}(\cdot)$ is the $sth$ component of $U_{2s}(\cdot)$,
$R_k^{ij}$ is the $(i,j)th$ element of $R_k^{-1}, k=1,\ldots,n;
i,j=1,\ldots, m_k$ and $X_{iks}^J$ is the $sth$ element of
$J_{\beta_0^{(r)}}^TX_k$. If $R_k, k=1,\ldots,n$ are not equal to each other,
the arguments contained in \cite{Wang-Xue-Zhu-Chong-2010} that show the term $U_{2s}(\beta_0^{(r)})$ is asymptotically negligible do not apply, and thus we cannot show the asymptotic normality
of $\tilde{\beta}$.
 Therefore, instead
of GEE (\ref{2.5}),}
 we incorporate bias correction which was previously used in \cite{Li-Zhu-Xue-Feng-2010}, leading to the bias-corrected GEE
\begin{equation}
\sum_{j=1}^n\hat{Z_j^0}^TR_j^{-1}(Y_j-\hat{\mathbf{g}}(\mathbb{X}_j^T\beta))=0,
\label{2.6}
\end{equation}
where
$$
\hat{Z}^0_j=\left(
      \begin{array}{c}
        \hat{g}'(X_{j1}^T\beta)(J_{\beta^{(r)}}^T(X_{j1}-\hat{E}[X_{j1}|X_{j1}^T\tilde{\beta}]))^T \\
        \vdots \\
        \hat{g}'(X_{jm_j}^T\beta)(J_{\beta^{(r)}}^T(X_{jm_j}-\hat{E}[X_{j1}|X_{jm_j}^T\tilde{\beta}]))^T \\
      \end{array}
    \right),
    j=1,\ldots,n,
$$
and $\hat{E}(X_{jk}|X_{jk}^T\tilde{\beta})$ is a nonparametric estimate of $E(X_{jk}|X_{jl}^T\beta_0)$ with $\beta_0$ replaced by the initial estimator $\tilde{\beta}$, that is
$$
\hat{E}(X_{jk}|X_{jk}^T{\beta})=\sum_{l_1=1}^n\sum_{l_2=1}^{m_{l_1}}W_{nl_1l_2}(X_{jk}^T{\beta},\beta)X_{l_1l_2}.
$$

In the following, and also in the proofs in the Appendix, with misuse of notation but for simplicity in writing,  we will write the matrix such as
$$
\left(\begin{array}{c}
        g'(X_{j1}^T\beta)(J_{\beta}^T(X_{j1}-E[X_{j1}|X_{j1}^T{\beta}]))^T \\
        \vdots \\
        g'(X_{jm_j}^T\beta)(J_{\beta}^T(X_{jm_j}-E[X_{jm_j}|X_{jm_j}^T{\beta}]))^T \\
      \end{array}
    \right)
$$
simply as
$\mathbf{g}'(\mathbb{X}_j^T\beta)J_{\beta}^T(\mathbb{X}_j-{E}(\mathbb{X}_j|\mathbb{X}_j^{{\top}}{\beta})$
and take $\mathbb{X}_j-E(\mathbb{X}_j|\mathbb{X}_j^{{\top}}\beta)$
to denote the $m_j\times p$ matrix with entries
$X_{jlq}-{E}[X_{jlq}|X_{jlq}^T{\beta}], 1\le l\le m_j, 1\le q\le p$.

Denote the solution of (\ref{2.6}) by
${\hat{\beta}^{(r)}_* }$ (the notations $\hat{\beta}$ and $\hat{\beta}^{(r)}$ are reserved for the estimator based on smooth-threshold generalized estimating equations later when we deal with variable selection), 
thus our final estimator for $\beta$ is
$\hat{\beta}_*=\beta({\hat{\beta}^{(r)}}_*)$. We have the following asymptotic property for $\hat{\beta}_*$.

\begin{theorem} Under the regularity conditions given in the Appendix, and suppose the initial estimator $\tilde{\beta}$ is $\sqrt{n}$-consistent, then there exists a solution $\hat{\beta}_*$ of (\ref{2.6}) inside the ball $\mathbf{B}=\{\|\beta-\beta_0\|\le C{n}^{-1/2}\}$ for $C$ sufficiently large. Furthermore,
$$
\sqrt{n}(\hat{\beta}_*-\beta_0)\stackrel{d}{\longrightarrow}N(0,\Sigma_{a}),
$$
where
$$
\Sigma_{a}=J_{{\beta}_0^{(r)}}V^{-1}\Omega V^{-1} J_{{\beta}_0^{(r)}}^T.
$$
The matrices $V$ and $\Omega$ are defined in condition C7 of
the Appendix.
\end{theorem}

\begin{remark} We have assumed that $R_i$ are prespecified and known in the above. However, from the proof, one easily sees that when $R_i$ is replaced by a consistent estimator $\hat{R}_i$, the theorem still holds. When $\Sigma_1=\cdots=\Sigma_n$, a consistent estimator of $\Sigma_i$ is
$\sum_{j=1}^n\hat{\varepsilon}_j\hat{\varepsilon}_j^T/n$,
where
$$
\hat{\varepsilon}_j=(Y_{j1}-\tilde{g}(X_{j1}^T\tilde{\beta}),\ldots,Y_{jm_j}-\tilde{g}(X_{jm_j}^T\tilde{\beta}))^T,
$$
with $\tilde{g}$ and $\tilde{\beta}$ obtained from the working independence assumption \citep{balan}. Alternatively, when $R_i, 1\le i\le n$ depend on some fixed parameter $\alpha$, moments-based method can be used to estimate $\alpha$ consistently, resulting in consistent estimator of $R_i$ \citep{Liang-Zeger-1986}.
\end{remark}

\begin{remark} When $R_1=\cdots=R_n=R$, it is not necessary to use bias-corrected GEE (\ref{2.6}). In particular, when using GEE (\ref{2.5}), Lemma A.7 in \cite{Wang-Xue-Zhu-Chong-2010} can be followed line by line (with the extra simplification that we are dealing with single-index models instead of partially linear single-index models in that paper) to show that $\hat{\beta}_*$ is asymptotically normal with covariance matrix
$\Sigma_b=J_{{\beta}_0^{(r)}}V_1^{-1}\Omega V_1^{-1}J_{{\beta}_0^{(r)}}^T$,
where
$$
V_1=\lim_{n\rightarrow
\infty}\frac{1}{n}\sum_{k=1}^nE\Big[(\mathbf{g}'({\mathbb{X}_k}^T{\beta}_0)J_{{\beta}_0^{(r)}}^T\mathbb{X}_k)^TR_k^{-1}(\mathbf{g}'({\mathbb{X}_k}^T{\beta}_0)J_{{\beta}_0^{(r)}}^T\mathbb{X}_k)\Big].
$$
It is obvious that $V_{1}\ge V$ (i.e. $V_1-V$ is nonnegative definite) and thus $\Sigma_b\le\Sigma_a$, which means estimator obtained from (\ref{2.5}) is more efficient than that obtained from (\ref{2.6}). However, theoretically, using bias-correction leads to simpler assumptions on the bandwidth. In particular, unlike the theoretical results presented in \cite{chang10}, we do not need to use different bandwidths when estimating $g$ and $g'$ if (\ref{2.6}) is used. In our simulation results, our experience is that empirically the difference between using (\ref{2.5}) and (\ref{2.6}) is very small and thus we only report the simulation results based on bias-corrected GEE only. When $R_i$ are not all equal, the original proof in \cite{Wang-Xue-Zhu-Chong-2010} fall through and this is the reason for proposing (\ref{2.6}) to make our presentation much more general and work in all cases.
\end{remark}

\section{Variable selection and the oracle property}
So far in our discussions, all the covariates are assumed to be important for predicting $Y$. However, in many practical situations, some covariate variables are independent of or have negligible correlations with the response variable. As mentioned in the introduction, many shrinkage based approaches have been proposed in the literature to solve this variable selection problem, most of which are based on penalty functions with a singularity at zero. As an alternative method, \cite{Ueki-2009} proposed smooth-threshold estimating equations (SEE). This method is easily implemented with Newton-Raphson type algorithms, which is almost the same as solving the original estimating equations under the full model.

Let $\mathbf{A}=\{1,2,\ldots,p\}$ be the index set for the components of $\beta$. We make the sparsity assumption that some components of $\beta_0$ are zeros and without loss of generality assume the first $p_0$ components are nonzero and let $\mathbf{A}_0=\{1,2,\ldots,p_0\}$, and thus $\mathbf{A}_0^c$ contains all the indices of the zero components. Following \cite{Ueki-2009}, we propose the following smooth-threshold generalized estimating equations (SGEE) for simultaneous variable selection and estimation,
\begin{equation}
(I_{p-1}-\hat{D})\sum_{j=1}^n\hat{Z^0_j}^T{R}_j^{-1}(Y_j-\hat{\mathbf{g}}(\mathbb{X}_j^T\beta))+\hat{D}\beta^{(r)}=0,
\label{2.7}
\end{equation}
where
$$
\hat{Z}^0_j=\left(
      \begin{array}{c}
        \hat{g}'(X_{j1}^T\beta)(J_{\beta^{(r)}}^T[X_{j1}-\hat{E}(X_{j1}|X_{j1}^T\tilde{\beta})])^T \\
        \vdots \\
        \hat{g}'(X_{jm_j}^T\beta)(J_{\beta^{(r)}}^T[X_{jm_j}-\hat{E}(X_{jm_j}|X_{jm_j}^T\tilde{\beta})])^T \\
      \end{array}
    \right),
    j=1,\ldots,n;
$$
$$
    \hat{D}=\left(
        \begin{array}{cccccc}
          \hat{\delta}_1 &  &  &  &  &  \\
           & \ddots &  &  & 0 &  \\
           &  & \hat{\delta}_{r-1} &  &  &  \\
           &  &  & \hat{\delta}_{r+1} &  &  \\
           & 0 &  &  & \ddots &  \\
           &  &  &  &  & \hat{\delta}_p \\
        \end{array}
      \right),
$$
with $\hat{\delta}_i=min(1,\frac{\lambda}{|\tilde{\beta}_i|^{1+\gamma}}),i=1,\ldots,p, i\neq r; $ and $\tilde{\beta}$ is the initial $\sqrt{n}$-consistent estimator as before. The estimate of $\beta$ obtained from SGEE (\ref{2.7}) is denoted by $\hat{\beta}$ and the set of estimated nonzero indices is $\hat{\mathbf{A}}=\{i:\hat{\beta}_i\neq 0\}$.

From (\ref{2.7}), we see that $\hat{\delta}_i=1$ implies $\hat{\beta}_i=0$, while if $\hat{\delta}_i$ is negligibly close to zero, then (\ref{2.7}) is similar to (\ref{2.6}). The choice $\hat{\delta}_i=min(1,\frac{\lambda}{|\tilde{\beta}_i|^{1+\gamma}})$, proposed in \cite{Ueki-2009}, satisfies the desired property that $\hat{\delta}_i=1$ for insignificant variables and negligible for significant variables, if the parameter $\lambda>0$ is appropriately chosen.

\begin{theorem} Suppose the conditions C1-C7 in the Appendix hold, and $r\le p_0$. For any positive $\lambda$ and $\gamma$ such that $n^{1/2}\lambda\rightarrow 0$ and $n^{(1+\gamma)/2}\lambda\rightarrow\infty$ as $n\rightarrow\infty$,  we have: (i) variable selection consistency, i.e. $P(\mathbf{\hat{A}}=\mathbf{A}_0)\rightarrow 1$; (ii) asymptotic normality, i.e. $n^{1/2}(\hat{\beta}_{\mathbf{A}_0}-\beta_{0,\mathbf{A}_0})$ is
asymptotically normal with mean zero and covariance matrix the same as when $\mathbf{A}_0$ is known.
\end{theorem}
We note that in the statement of the theorem, we need to assume $r\le p_0$, that is, the removed component is significant. In practice, we select this component based on the initial estimator under the full model, and choose the component that has the largest absolute value.

To use the SGEE in practice, we need to choose appropriately the tuning parameters $(\lambda,\gamma)$. Following \cite{Ueki-2009}, we use BIC-type criterion to choose these two parameters. That is, we choose $ (\lambda,\gamma)$ as the minimizer of
\begin{equation*}
BIC_{\lambda.\gamma}=\sum_{i=1}^n(Y_i-\hat{\mathbf{g}}(\mathbb{X}_i^T\hat{\beta}_{\lambda,\gamma},\hat{\beta}_{\lambda,\gamma}))^TR_i^{-1}(Y_i-\hat{\mathbf{g}}(\mathbb{X}_i^T\hat{\beta}_{\lambda,\gamma},\hat{\beta}_{\lambda,\gamma}))+df_{\lambda,\gamma}\log(n),
\label{2.8}
\end{equation*}
where $\hat{\beta}_{\lambda,\gamma}$ is the estimator for given $(\lambda, \gamma)$, $df_{\lambda,\gamma}$ is the
number of estimated nonzero parameters.

\section{Numerical studies}
\subsection{Simulations}
In this section, we carry out some simulations to evaluate the
finite sample performance of our proposed method. 
\textcolor{black}{
For each example below, we generate 200 data sets, each consisting
of $n=50$ or $100$ subjects. For Examples 1-3, we have $m_k\equiv m=3$
observations per subject. Within a cluster, the covariance of the
error is specified by
$Cov(\varepsilon_{m'},\varepsilon_{m''})=0.5^{|m'-m''|},
m',m''=1,\ldots,m$. For Example 4, we have $m_k=1,2,3$ for
$k\leq n/3$, $n/3<k\leq 2n/3$, and $k>2n/3$, respectively. Within a cluster, the covariance of the error is
specified by $Cov(\varepsilon_{km'},\varepsilon_{km''})=0.5^{|m'-m'|}, 1\le m',m''\le m_k$.}
The kernel function is taken to be
$K(x)=\frac{3}{4}(1-x^2)$ if $|x|\leq1$, 0 otherwise, and the
bandwidth $h$ is selected by leave-one-out cross validation. We
compare the proposed estimator $\hat{\beta}$ with the oracle
estimator (when the zero coefficients are known), the estimator
$\hat{\beta}_*$, and also with $\hat{\beta}_I$, which is the
solution of SGEE (\ref{2.7}) using identity matrices as working covariance
matrices. The following criterions are considered.
\begin{itemize}
  \item The square of the R statistic: $R^2=\frac{|\hat{\beta}^T\beta_0|^2}{|\beta_0^T\beta_0|^2}$;
  \item The number of zero coefficients and nonzero coefficients obtained by different methods: ``TN'' is the average number of zero coefficients correctly estimated as zero, and ``TP'' is the number of nonzero coefficients identified as nonzero.
\end{itemize}

In these simulations, for SGEE estimator, the common intra-cluster covariance matrix is estimated nonparametrically from the residuals based on the initial estimator assuming working independence.

Example 1. Consider the single-index model for longitudinal data
\begin{equation*}
Y_{ij}=exp(X_{ij}^T\beta_0)+\varepsilon_{ij}, \quad i=1,\ldots,n,
j=1,\ldots,3,\label{3.1}
\end{equation*}
where $X_{ij}=(X_{ij1},\ldots,X_{ij6})^T$ was generated from multivariate normal distribution with identity covariance matrix. The true parameter is $\beta_0=\frac{1}{\sqrt{2}}(1,1,0,0,0,0)^T$. The numerical results
are reported in Table 1.

Example 2. Similarly to Example 1 except that we let $\beta_0=\frac{1}{\sqrt{1.4}}(1,0.6,0.2,0,0,0)^T$. The
numerical results are reported in Table 2.

Example 3. Similar to Example 1, except we use a different link function $g(X^T\beta_0)=\sin(X^T\beta_0)$ which is nonmonotone. The numerical results are reported in Table 3.

\textcolor{black}{ Example 4. Similar to Example 1, except that
 $m_k, k=1,\ldots,n$, are different. The numerical results are reported in Table
4.}

Tables 1-4 show that for our three examples, SGEE can satisfactorily identify the true model. Besides, it is advantageous to take into account the correlation of the observations.

\begin{table}
\centering\begin{tabular}{ccccc}
  \hline
  n  & Method & $R^2$  &  TN   & TP   \\ \hline
     & Oracle & 0.9982 &  4     & 2    \\
     & $\hat{\beta}_*$ & 0.9895 &  0     & 2    \\
  50 & $\hat{\beta}$ & 0.9935 & 3.955  & 2    \\
     & $\hat{\beta}_I$ & 0.9817      & 3.525   & 2    \\
     &        &        &                  &      \\
     & Oracle & 0.9994 &  4     & 2    \\
     & $\hat{\beta}_*$ & 0.9962 &  0     & 2    \\
  100& $\hat{\beta}$ & 0.9960      & 3.985 & 2  \\
     & $\hat{\beta}_I$ & 0.9854      & 3.75 & 2  \\
  \hline
\end{tabular}
\caption{ Simulation results for Example 1.}
\end{table}

\begin{table}
\centering
\begin{tabular}{ccccc}
  \hline
  n  & Method & $R^2$  &  TN   & TP  \\ \hline
     & Oracle & 0.9970 &  3     & 3    \\
     & $\hat{\beta}_*$ & 0.9840 &  0     & 3    \\
  50 &$\hat{\beta}$ & 0.9329      & 2.78  & 2.255\\
     &$\hat{\beta}_I$ & 0.9128      & 2.615 & 2.47 \\
     &        &        &                 &      \\
     & Oracle & 0.9986      & 3     & 3    \\
     & $\hat{\beta}_*$ & 0.9925 &  0     & 3    \\
  100& $\hat{\beta}$ & 0.9527      & 2.94  & 2.3  \\
     & $\hat{\beta}_I$ & 0.9411     & 2.8   & 2.395 \\
  \hline
\end{tabular}
\caption{ Simulation results  for Example 2.}
\end{table}

\begin{table}
\centering\begin{tabular}{ccccc}
  \hline
  n  & Method & $R^2$  &  TN   & TP  \\ \hline
     & Oracle & 0.9832   & 4     & 2    \\
     & $\hat{\beta}_*$ & 0.9171 &  0     & 2    \\
  50 & $\hat{\beta}$ & 0.9756     & 3.885 & 2    \\
     & $\hat{\beta}_I$& 0.9558      & 3.775 & 2    \\
     &        &        &                  &      \\
     & Oracle & 0.9928      & 4     & 2    \\
     & $\hat{\beta}_*$ & 0.9648 &  0     & 2    \\
  100& $\hat{\beta}$ & 0.9912      & 3.94  & 2 \\
     & $\hat{\beta}_I$& 0.9889      & 3.88 & 2 \\
  \hline
\end{tabular}
\caption{Simulation results for Example 3.}
\end{table}

\begin{table}
\centering\begin{tabular}{ccccc}
  \hline
  n  & Method & $R^2$  &  TN   & TP   \\ \hline
     & Oracle & 0.9980 &  4     & 2    \\
     & $\hat{\beta}_*$ & 0.9806 &  0     & 2    \\
  50 & $\hat{\beta}$ & 0.9972 & 3.965  & 2    \\
     & $\hat{\beta}_I$ & 0.9966      & 3.85   & 2    \\
     &        &        &                  &      \\
     & Oracle & 0.9996 &  4     & 2    \\
     & $\hat{\beta}_*$ & 0.9931 &  0     & 2    \\
  100& $\hat{\beta}$ & 0.9990      & 3.995 & 2  \\
     & $\hat{\beta}_I$ & 0.9986      & 3.99 & 2  \\
  \hline
\end{tabular}
\caption{\textcolor{black}{ Simulation results for Example 4.}}
\end{table}

\subsection{Real data}

We now apply the proposed procedure to the CD4 data from the
Multi-Center AIDS Cohort Study. This data set has been studied in
\cite{Kaslow-Ostrow-Detels-Phair-Plok-Rinaldo-1987, Fan-Li-2004,
Fan-Huang-Li-2007, Li-Zhu-Xue-Feng-2010}. The data set contains the
human immunodeficiency virus (HIV) status of 283 homosexual men who
were infected with HIV during the follow-up period between 1984 and
1991. Details of the study design, methods, and medical implications
can be found in \cite{Kaslow-Ostrow-Detels-Phair-Plok-Rinaldo-1987}.
All individuals were scheduled to have their measurements made
during semiannual visits. However, many participants missed some of
their scheduled visits resulting in different measurement time
points and unequal number of measurements per individual. In our
analysis, we let $y_{ij}$ be the CD4 cell counts for individual $i$
at the $j$th visit, $x_{ij1}$ be the smoking status with 1 for a
smoker and 0 for a nonsmoker, $x_{ij2}$ be the person's age, and
$x_{ij3}$ be last measured CD4 level before HIV infection. For
exploratory purposes, we also consider possible interactions of the
covariates and also squares of $x_{ij2}$ and $x_{ij3}$, resulting in
the following model:
\begin{eqnarray*}
y_{ij}&=&g\big(x_{ij1}\beta_1+x_{ij2}\beta_2+x_{ij3}\beta_3+x_{ij2}^2\beta_4+x_{ij3}^2\beta_5\\
&+&x_{ij1}x_{ij2}\beta_6
+x_{ij1}x_{ij3}\beta_7+x_{ij2}x_{ij3}\beta_8\big)+\varepsilon_{ij}.
\end{eqnarray*}

We apply the SGEE approach to this data set to select significant
variables and estimate the effects. The tuning parameters $\lambda$
and $\gamma$ are selected by the BIC-type criterion. For any
individual, we assume the  correlation between visits at time
$t_{j_1}$ and $t_{j_2}$ is $\alpha^{|t_{j_1}-t_{j_2}|}$. The fitted
model is
$$
y_{ij}\sim g(0.4531x_{ij1}-0.6744x_{ij2}+0.5829x_{ij3})
$$
By our variable selection procedure, we can see that only the linear terms are significant.

\begin{appendix}
\begin{center}
 \Large \sc{Appendix}
%\vskip 0.3 cm
%\normalsize \textit{Assumptions}
\end{center}
\vspace{0.3 cm}
\appendix

In order to study the asymptotic behavior of the estimator, the following standard assumptions are imposed \citep{Li-Zhu-Xue-Feng-2010}.
\begin{itemize}
  \item C1. The density function $f_{ij}(t)$ of $X_{ij}^T\beta$ is bounded away from zero and continuously differentiable on $\{t:t=X_{ij}^T\beta, X_{ij}\in A, i=1,\ldots,n;j=1,\ldots,m_i\}$ and $A$ is the support of $X_{ij}$ which is assumed to be compact.
  \item C2. The function $g(\cdot)$ is twice continuously differentiable, and $E(X_{klq}|X_{klq}^T\beta=x), 1\le l\le m_k, 1\le k\le n, 1\le q\le p$ as a function of $x$ is Lipschitz continuous.
  \item C3. The kernel $K$ is a bounded, continuous and symmetric
  probability density function, satisfying
  $$\int_{-\infty}^{\infty}u^2K(u)du<\infty.$$
  \item C4. There exists a positive constant $M$, such that $\max_{1\leq k\leq n, 1\leq j\leq m_k}E(\varepsilon^4_{kj})\leq M<\infty$.
  \item C5. The bandwidth $h$ satisfies $nh^3\rightarrow\infty, nh^8\rightarrow0$.
  \item C6. The eigenvalues of $R_i$ and $\Sigma_i$ are uniformly bounded and bounded away from zero.
  \item C7. $\Omega=\lim_{n\rightarrow\infty}\frac{1}{n}\sum_{k=1}^n E\Big\{\big(\mathbf{g}'(\mathbb{X}_k^T{\beta}_0)J_{{\beta}_0^{(r)}}^T \tilde{\mathbb{X}}_k^0\big)^TR_k^{-1}\varepsilon_k\Big\}^{\otimes2}$ is positive definite, where we use the notation $\tilde{\mathbb{X}}_k^0=\mathbb{X}_k-E(\mathbb{X}_k|\mathbb{X}_k^T\beta_0)$ is $m_k\times p$ matrix with entries $X_{klq}-{E}[X_{klq}|X_{klq}^T{\beta}], 1\le l\le m_k, 1\le q\le p$, and $E(A)^{\otimes2}=E(AA^T)$ for any matrix $A$.\\
 $V=\lim_{n\rightarrow\infty}\frac{1}{n}\sum_{k=1}^nE\Big[(\mathbf{g}'({\mathbb{X}_k}^T{\beta}_0)J_{{\beta}_0^{(r)}}^T\tilde{\mathbb{X}}^0_k)^TR_k^{-1}(\mathbf{g}'({\mathbb{X}_k}^T{\beta}_0)J_{{\beta}_0^{(r)}}^T\tilde{\mathbb{X}}^0_k)\Big]$ is also positive definite.
\end{itemize}
\textcolor{black}{{\bf Remark.} Note that in condition C1 we allow the distributions of $X_{ij}^T\beta$ to be different for different $i,j$, and in particular $m_i, i=1,\ldots, n$, are
not required to be the same. }

\textit{Proof of Theorem 1.} Proof of existence of
$\sqrt{n}$-consistent solution to (\ref{2.6}) is almost same as in
\cite{Wang-Xue-Zhu-Chong-2010} and omitted here. Thus we proceed to
consider asymptotic normality. By (\ref{2.6}), since
\begin{eqnarray*}
\sum_{k=1}^n\hat{{Z}^0_k}^T{R}_k^{-1}(Y_k-\hat{\mathbf{g}}(\mathbb{X}_k^T\hat{\beta}_*))=0,
\end{eqnarray*}
it follows
\begin{eqnarray*}
&&\sum_{k=1}^n\hat{{Z^0_k}}^T{R}_k^{-1}(Y_k-\hat{\mathbf{g}}(\mathbb{X}_k^T\hat{\beta}_*))\\
&=&\sum_{k=1}^n\big(\mathbf{g}'(\mathbb{X}_k^T{\beta}_0)J_{\hat{\beta}_*^{(r)}}^T[\mathbb{X}_k-E(\mathbb{X}_k|\mathbb{X}_k^T{\beta}_0)]\big)^T
R_k^{-1}\varepsilon_k\\
&+&\sum_{k=1}^n\big([\hat{\mathbf{g}}'(\mathbb{X}_k^T\hat{\beta}_*)-\mathbf{g}'(\mathbb{X}_k^T{\beta}_0)]J_{\hat{\beta}_*^{(r)}}^T[\mathbb{X}_k-\hat{E}(\mathbb{X}_k|\mathbb{X}_k^T\tilde{\beta})]\big)^T
R_k^{-1}\varepsilon_k\\
&+&\sum_{k=1}^n\big(\mathbf{g}'(\mathbb{X}_k^T{\beta}_0)J_{\hat{\beta}_*^{(r)}}^T[E(\mathbb{X}_k|\mathbb{X}_k^T{\beta}_0)-\hat{E}(\mathbb{X}_k|\mathbb{X}_k^T\tilde{\beta})]\big)^T
R_k^{-1}\varepsilon_k\\
&+&\sum_{k=1}^n\big(\mathbf{g}'(\mathbb{X}_k^T{\beta}_0)J_{\hat{\beta}_*^{(r)}}^T[\mathbb{X}_k-\hat{E}(\mathbb{X}_k|\mathbb{X}_k^T\tilde{\beta})]\big)^TR_k^{-1}
\big(\mathbf{g}(\mathbb{X}_k^T{\beta}_0)-\hat{\mathbf{g}}(\mathbb{X}_k^T\hat{\beta}_*)\big)\\
&+&\sum_{k=1}^n\big([\hat{\mathbf{g}}'(\mathbb{X}_k^T\hat{\beta}_*)-\mathbf{g}'(\mathbb{X}_k^T{\beta}_0)]J_{\hat{\beta}_*^{(r)}}^T[\mathbb{X}_k-\hat{E}(\mathbb{X}_k|\mathbb{X}_k^T\tilde{\beta})]\big)^T
R_k^{-1}\big(\mathbf{g}(\mathbb{X}_k^T{\beta}_0)-\hat{\mathbf{g}}(\mathbb{X}_k^T\hat{\beta}_*)\big)\\
&:=&Q_1(\hat{\beta}_*^{(r)})+Q_2(\hat{\beta}_*^{(r)})+Q_3(\hat{\beta}_*^{(r)})+Q_4(\hat{\beta}_*^{(r)})+Q_5(\hat{\beta}_*^{(r)}).
\hspace{35mm} (A.1)
\end{eqnarray*}
Noting that
$J_{{\hat{\beta}}_*^{(r)}}-J_{{\beta}_0^{(r)}}=O_p(n^{-1/2})$, we have
$$
Q_1({\hat{\beta}}_*^{(r)})-U({\beta}_0^{(r)})=o_p(\sqrt{n}), \eqno(A.2)
$$
where
$$
U({\beta}_0^{(r)})=\sum_{k=1}^n\big(\mathbf{g}'(\mathbb{X}_k^T{\beta}_0)J_{{\beta}_0^{(r)}}^T(\mathbb{X}_k-E[\mathbb{X}_k|\mathbb{X}_k^T{\beta}_0])\big)^TR_k^{-1}\varepsilon_k.
$$
For $Q_2({\hat{\beta}}_*^{(r)})$, denote
$$
R_k^{-1}=\left(
         \begin{array}{ccc}
           R_{k11}^{-1} & \cdots & R_{k1m_k}^{-1} \\
           \vdots & \ddots & \vdots \\
           R_{km_k1}^{-1} & \cdots & R_{km_km_k}^{-1} \\
         \end{array}
       \right),
$$ then
$$
Q_2({\hat{\beta}}_*^{(r)})=\sum_{k=1}^n\sum_{j=1}^{m_k}\varepsilon_{kj}\sum_{i=1}^{m_k}R_{kij}^{-1}[\hat{g}'({X}_{ki}^T\hat{\beta}_*)-g'({X}_{ki}^T{\beta}_0)]J_{{\hat{\beta}}_*^{(r)}}^T[{X}_{ki}-\hat{E}({X}_{ki}|{X}_k^T\tilde{\beta})].
\eqno(A.3)
$$
 
\textcolor{black}{
 Note that $\hat{\beta}_*, \tilde{\beta}\in\mathbf{B}$, together with conditions C2 and C3, we have
 \begin{eqnarray*}
Q_2({\hat{\beta}}_*^{(r)})&=&\sum_{k=1}^n\sum_{j=1}^{m_k}\varepsilon_{kj}\sum_{i=1}^{m_k}
R_{kij}^{-1}[\hat{g}'({X}_{ki}^T\beta_0)-g'({X}_{ki}^T{\beta}_0)+\frac{\partial}{\partial
\beta^{(r)}}\hat{g}'({X}_{ki}^T\bar{\beta}_1)(\hat{\beta}_*^{(r)}-\beta_0^{(r)})]\\
&\times&J_{{\hat{\beta}}_*^{(r)}}^T[{X}_{ki}-\hat{E}({X}_{ki}|{X}_k^T{\beta}_0)+\frac{\partial}{\partial
\beta^{(r)}}\hat{E}({X}_{ki}|{X}_k^T\bar{\beta}_2)(\hat{\beta}_*^{(r)}-\beta_0^{(r)})],
 \end{eqnarray*}
 where $\bar{\beta}_1$ and $\bar{\beta}_2$ are the intermediate
 values between $\beta_0$ and $\hat{\beta}_*$. Thus,
$$
Q_2({\hat{\beta}}_*^{(r)})=\sum_{k=1}^n\sum_{j=1}^{m_k}\varepsilon_{kj}\sum_{i=1}^{m_k}R_{kij}^{-1}
[\hat{g}'({X}_{ki}^T{\beta}_0)-g'({X}_{ki}^T{\beta}_0)]J_{{\beta}_0^{(r)}}^T[{X}_{ki}-\hat{E}({X}_{ki}|{X}_k^T{\beta}_0)]+o_p(\sqrt{n}).
$$
}

\textcolor{black}{ Let
$Q_2({\hat{\beta}}_*^{(r)})=J_{{\beta}_0^{(r)}}^TQ_2({\hat{\beta}}_*^{(r)})^*$,
where the $sth$ component of $Q_2({\hat{\beta}}_*^{(r)})^*$ is
$$
Q_2({\hat{\beta}}_*^{(r)})_s^*=\sum_{k=1}^n\sum_{j=1}^{m_k}\sum_{i=1}^{m_k}\varepsilon_{kj}R_{kij}^{-1}
[\hat{g}'({X}_{ki}^T{\beta}_0)-g'({X}_{ki}^T{\beta}_0)][{X}_{kis}-\hat{E}({X}_{kis}|{X}_{ki}^T{\beta}_0)].
$$
By (\ref{2.4}), let
$\tilde{X}_{kis}=[{X}_{kis}-\hat{E}({X}_{kis}|{X}_{ki}^T{\beta}_0)]$
be the $sth$ component of
${X}_{ki}-\hat{E}({X}_{ki}|{X}_{ki}^T{\beta}_0)$, we have
\begin{eqnarray*}
&&Q_2({\hat{\beta}}_*^{(r)})_s^*\\
&=&\sum_{k=1}^n\sum_{i=1}^{m_k}\sum_{j=1}^{m_k}\varepsilon_{kj}\tilde{X}_{kis}R_{kij}^{-1}
\Big[\sum_{l_1=1}^n\sum_{l_2=1}^{m_{l_1}}\tilde{W}_{nl_1l_2}({X}_{ki}^T{\beta}_0,{\beta}_0)
g({X}_{ki}^T{\beta}_0)-g'({X}_{ki}^T{\beta}_0)\Big]\\
&+&\sum_{k=1}^n\sum_{j=1}^{m_k}
\tilde{W}_{nkj}({X}_{kj}^T{\beta}_0,{\beta}_0)\tilde{X}_{kjs}\varepsilon_{kj}^2R_{kjj}^{-1}
+\sum_{k=1}^n\sum_{j\neq i}^{m_k}
\tilde{W}_{nkj}({X}_{ki}^T{\beta}_0,{\beta}_0)\tilde{X}_{kis}\varepsilon_{kj}^2R_{kij}^{-1}\\
&+&\sum_{k=1}^n\sum_{i=1}^{m_k}\sum_{j=1}^{m_k}\sum_{l_1\neq
k}^n\sum_{l_2\neq
i}^{m_{l_1}}\tilde{W}_{nl_1l_2}({X}_{ki}^T{\beta}_0,{\beta}_0)\tilde{X}_{kis}R_{kij}^{-1}\varepsilon_{kj}\varepsilon_{l_1l_2}\\
&:=&Q_{21s}^*+Q_{22s}^*+Q_{23s}^*+Q_{24s}^*.
\end{eqnarray*}
}

Similar to the proof of Lemma A.4 in \cite{Li-Zhu-Xue-Feng-2010}, utilizing also Lemmas A.1-A.3 there, we can show that
$Q_2(\hat{\beta}_*^{(r)})_s^*=o_p(\sqrt{n})$ and thus
$$
Q_2(\hat{\beta}_*^{(r)})=o_p(\sqrt{n}).\eqno(A.4)
$$
Similarly, we can obtain
$$Q_3({\hat{\beta}}_*^{(r)})=o_p(\sqrt{n}),\quad Q_5({\hat{\beta}}_*^{(r)})=o_p(\sqrt{n}).\eqno(A.5)$$
For $Q_4(\hat{\beta}_*^{(r)})$, simple calculations yield
\begin{eqnarray*}
Q_4(\hat{\beta}_*^{(r)})&=&
\sum_{k=1}^n\sum_{i=1}^{m_k}\sum_{j=1}^{m_k}
g'({X}_{ki}^T\hat{\beta}_*)J_{\hat{\beta}_*^{(r)}}^T[{X}_{ki}-\hat{E}({X}_{ki}|{X}_{ki}^T\hat{\beta}_*)]R_{kij}^{-1}
\big({g}({X}_{kj}^T\hat{\beta}_*)-\hat{g}({X}_{kj}^T\hat{\beta}_*)\big)\\
&-&\sum_{k=1}^n\sum_{i=1}^{m_k}\sum_{j=1}^{m_k}
g'({X}_{ki}^T{\beta}_0)J_{{\beta}_0^{(r)}}^T[{X}_{ki}-{E}({X}_{ki}|{X}_{ki}^T{\beta}_0)]R_{kij}^{-1}\\
&&\times
g'(X_{kj}^T\beta_0)\big\{J_{{\beta}_0^{(r)}}^T[{X}_{ki}-{E}({X}_{ki}|{X}_{ki}^T{\beta}_0)]\big\}^T({\hat{\beta}}_*^{(r)}-{\beta}_0^{(r)})
+o_p(\sqrt{n})\\
&=&Q_{41}(\hat{\beta}_*^{(r)})+Q_{42}(\hat{\beta}_*^{(r)})+o_p(\sqrt{n}).
\end{eqnarray*}
It is easy to show that $Q_{41}(\hat{\beta}_*^{(r)})=o_p(\sqrt{n})$ and that
$$
Q_{42}({\hat{\beta}}_*^{(r)})-nV({\hat{\beta}}_*^{(r)}-{\beta}_0^{(r)})=o_p(\sqrt{n}),\eqno(A.6)
$$
where
$$
V=\lim_n\frac{1}{n}\sum_{k=1}^nE\Big[(\mathbf{g}'(\mathbb{X}_k^T{\beta}_0)J_{{\beta}_0^{(r)}}^T\tilde{\mathbb{X}}_{k}^0)^TR_k^{-1}
(\mathbf{g}'({\mathbb{X}_k}^T{\beta}_0)J_{{\beta}_0^{(r)}}^T\tilde{\mathbb{X}}_{k}^0)\Big],
\eqno(A.7)
$$
is a positive definite matrix. Thus
$$
Q_{4}({\hat{\beta}}_*^{(r)})-nV({\hat{\beta}}_*^{(r)}-{\beta}_0^{(r)})=o_p(\sqrt{n}),\eqno(A.8)
$$

In summary, by estimating equation (\ref{2.6}), together with (A.2),
(A.4), (A.5) and (A.8), it follows
\begin{eqnarray*}
\hspace{20mm}0&=&
Q_1({\hat{\beta}}_*^{(r)})+Q_2({\hat{\beta}}_*^{(r)})+Q_3({\hat{\beta}}_*^{(r)})+Q_4({\hat{\beta}}_*^{(r)})+Q_5({\hat{\beta}}_*^{(r)})\\
&=&U({\beta}_0^{(r)})+o_p(\sqrt{n})-nV(\hat{\beta}_*^{(r)}-{\beta}_0^{(r)})\\
&\Rightarrow&\sqrt{n}({\hat{\beta}}_*^{(r)}-{\beta}_0^{(r)})=V^{-1}n^{-1/2}U({\beta}_0^{(r)})+o_p(1).\hspace{32mm}(A.9)
\end{eqnarray*}
Thus, we have
$$
\sqrt{n}(\hat{\beta}_*-{\beta}_0)=J_{{\beta}_0^{(r)}}V^{-1}n^{-\frac{1}{2}}U({\beta}_0^{(r)})+o_p(1).\eqno(A.10)
$$
The asymptotic normality of $\hat{\beta}_*$ directly follows from this representation and the central limit theorem.
$\square$

\textit{Proof of Theorem 2}. First, for $j\in \mathbf{A}_{0}^c$, we have $|\tilde{\beta}_j^{(r)}|=O({n}^{-1/2})$ by the assumption of $\sqrt{n}$-consistency of the initial estimator. Using the condition on $\lambda$ in the statement of the theorem, we get
$$
P(\lambda/|\tilde{\beta}_j^{(r)}|^{1+\gamma}<1)\rightarrow 0, j\in\mathbf{A}_{0}^c, \eqno(A.12)
$$
and thus
$$P(\hat{\delta}_j=1\ for\ all\ j\in\mathbf{A}_0^c)\rightarrow 1.$$
On the other hand, we have for any $\epsilon>0$ and $j\in\mathbf{A}_{0}-\{r\}$,
$$
P(\hat{\delta}_j>n^{-1/2}\epsilon)=P(\lambda n^{1/2}/\epsilon>|\tilde{\beta}_j^{(r)}|^{1+\gamma})\rightarrow 0
$$
using that $\lambda n^{1/2}\rightarrow 0$ and that
$|\tilde{\beta}_j^{(r)}|$ is bounded away from zero. Thus
$\hat{\delta}_j=o_p(n^{-1/2})$ for each $j\in\mathbf{A}_{0}-\{r\}$,
implying trivially
 $P(\hat{\delta}_j<1\ for\ all\ j\in\mathbf{A}_{0}-\{r\})\rightarrow 1$, and (i) is proved.

Next, we prove (ii). From (i) and the assumption that the $rth$ component of $\beta_0$ is nonzero, the SGEE coincide with
$$(1-\hat{\delta}_j)u_j(\hat{\beta}^{(r)})+\hat{\delta}_j\hat{\beta}_j^{(r)}=0,\ for\ j\in\mathbf{A}_{0}-\{r\} \eqno(A.13)$$
and $\hat{\beta}_j=0$ for $j\in\mathbf{A}_0^c$, with probability
tending to one, where $u_j(\hat{\beta}^{(r)})$ is the $jth$
component of
$\sum_{k=1}^n\hat{Z^0_k}^T{R}_k^{-1}(Y_k-\hat{\mathbf{g}}(\mathbb{X}_k^T\hat{\beta}))$,
$j\in\mathbf{A}_{0}-\{r\}$. Using that
$\hat{\delta}_j=o_p(n^{-1/2})$ for $j\in\mathbf{A}_{0}-\{r\}$, it is
easy to show that (A.13) is asymptotically equivalent to
$u_j(\hat{\beta}^{(r)})=0$ and the asymptotic normality follows the
same way as in the proof of Theorem 1.

\end{appendix}

\subsection*{Acknowledgements}
We thank an Associate Editor and two reviewers for their careful reading of the manuscript and helpful comments that led to an improvement of the manuscript.
Qihua Wang's research was supported by the  National Science Fund for Distinguished Young Scholars in China (10725106), the National Natural Science Foundation of China (10671198),  the National Science Fund for Creative Research Groups in China and a grant from Key Lab of Random Complex Structures and Data Science, Chinese Academy of Science. Heng Lian's research was supported by Singapore Ministry of Education Tier 1 RG36/09.

\bibliographystyle{elsart-harv}
\bibliography{reference}

\end{document}